# Two-Dimensional Wide Dynamic Range Displacement Sensor using Dielectric Resonator Coupled Microwave Circuit

Premsai Regalla and A.V. Praveen Kumar

*Abstract*— In this paper, the authors propose a two-dimensional (2D), wide dynamic range, linear displacement sensor using microwave methods. The microwave sensor circuit employs a cylindrical dielectric resonator (DR) proximity coupled to a pair of orthogonal microstrip lines formed on a microwave substrate. The DR rests on the substrate and is free to be displaced between the strips on the 2D plane of the substrate. The strips excite the particular resonant mode of the DR, the intensity of which varies with the DR's proximity to the strips. The DR's position can thus be read out in terms of the 2-port S-parameters of the circuit, at a fixed frequency determined by the resonant mode of DR. Such fixed frequency sensors are robust in operation and cost-effective in realization, an important aspect of this sensor. Initial one-dimensional (1D) positioning simulations (ANSYS HFSS tool) of the sensor through three fixed representative paths on the substrate (i.e., horizontal, vertical and diagonal) reveal that the S-parameters vary monotonically with the displacement. Prototype measurements reveal a dynamic range of 23 mm for horizontal/ vertical displacement, and 30 mm for diagonal (*dr*) displacement at the resonant frequency of 3.67 GHz. Next, 2D positioning test is conducted and a technique for one-to-one mapping from the S-parameters to the 2D position is demonstrated. To conclude, the proposed sensor's performance is compared with that of existing 2D sensors.

## I. INTRODUCTION

Radiofrequency (RF) and microwave sensors are attractive candidates for various sensing applications. Sensors using RF/microwave methods are known mainly for their higher accuracy than the low-frequency methods (ultrasonic, capacitive, inductive etc.), and low-cost implementation compared to optical methods (laser interferometry) [1]-[3]. Microwave sensors utilize specialized microwave elements that are sensitive to the physical parameter under detection, leading to changes in their electrical characteristics such as the resonant frequency, bandwidth, magnitude response, and/or phase response. As a result, microwave sensors are widely proposed for applications such as the measurement of material properties [4],[5], solvent concentration [6], humidity [7], strain [8], pressure [9], surface resistance [10], finger-touch [11], etc. Recently, applications of these sensors have been extended to industrial automation [12] and aerospace engineering [13] in the form of displacement sensors [12]-[35]. Such sensors are mostly based on microwave elements like the split ring resonators (SRR) [12]-[19], [33]-[35], defected ground structures (DGS) [21], frequency selected structures (FSS) [22],[23], metamaterial inspired resonators [27]-[32], dielectric resonators [38]-[45] or bi-path transverse filtering sections [36], [37].

Displacement sensors are categorized as one-dimensional linear [14]-[17], two-dimensional linear [18]-[24], angular [25]-[33], [36],[37], and linear-cum-angular [34], [35] sensors. Two-dimensional (2D) linear displacement sensors [18]-[24] are highly applicable as alignment and position sensors in industrial environments. Based on the operating principle, these can further be classified as variable frequency sensors [16], [19],[21]-[24], [27]-[29], frequency splitting sensors [16], [18], variable bandwidth sensors [28],[36],[37], single frequency-variable magnitude sensors [14],[15],[20],[26], and single frequency-variable phase sensors [17],[30]. Among these, the sensors which work on variable frequency, bandwidth, or phase provide a wider dynamic range. However, these require an expensive vector network analyzer (VNA) based measurement for reading the changes in the respective output parameters of the sensor. On the other hand, fixed frequency-variable magnitude sensors provide the unique advantage of needing only a minimal setup comprising a narrow-band RF generator (oscillator) and a power meter attached to the sensor [39],[40]. Sensors which work on the above principle are highly accurate for their insensitivity to changes in frequency /phase caused by the changes in external parameters such as temperature, humidity, proximity effects, wear and tear of components, etc. Additionally, the choice of scattering parameters (S-parameters) as the output quantity of the sensor further improves the accuracy in the presence of input power level fluctuations [38]. As the only fixed frequency-variable magnitude 2D linear displacement sensor reported to date is [20], there is a need for further research in this direction.

It can be learned from the literature that the majority of microwave sensors are based on SRR or its derivatives. The dielectric resonator (DR) as a sensor element is a very recent innovation in the field of microwave sensors [38]-[45]. DR-based displacement sensors offer a unique sensor solution with compactness, wide dynamic range, and performance tunability [38]-[41]. To the best of the authors' knowledge,

This work was supported by the Department of Science and Technology (DST), Govt. of India, through the Science and Engineering Research Board (SERB) grant Ref. EMR/2017/001126, and partly by the DST-FIST grant Ref: SR/FST/ETI-346/2013.
The authors are with the Department of Electrical and Electronics Engineering, Birla Institute of Technology and Science, BITS Pilani, Pilani Campus, Rajasthan, India (e-mail: praveen.kumar@pilani.bits-pilani.ac.in, premsairegalla999@gmail.com).

DR based 2D displacement sensors are also not reported before.

In view of the above, in the present paper, the authors discuss the design and analysis of a 2D displacement sensor which evolved from the previously reported 1D displacement sensor [39]. Although both these sensor designs work on fixed frequency-variable magnitude principle and employ identical circuit components, the present sensor provides 2D displacement sensing with much wider dynamic range than reported before.

The rest of the article is organized as follows: The sensor principle and design is presented in Section II. The sensitivity analysis of the sensor is carried out in Section III. Sensor measurements are presented in Section IV. Regression and error analysis is discussed in Section V and full 2D positioning is presented in Section VI. A parametric analysis of the DR and the sensor performance is presented in Section VII, and with a brief account of a practical application scenario, the paper is concluded in Section VIII.

## II. OPERATING PRINCIPLE AND DESIGN OF THE 2D SENSOR

The operating principle of the previously reported DR-based 1D displacement sensor [39] is briefly covered below with the help of its reproduced schematic diagram and response in Fig.1.

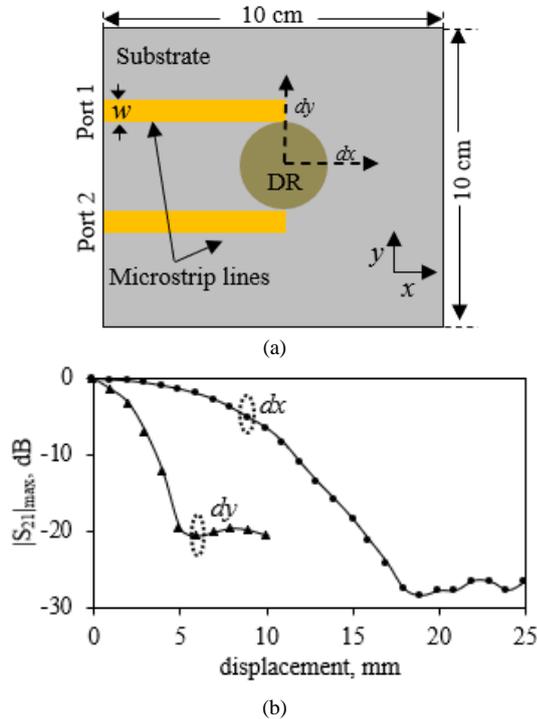

It comprises a microwave substrate on which a pair of parallel microstrip transmission lines are etched and placed between the lines is a cylindrical DR which is movable in close contact with the substrate. The substrate is RT/duroid with dielectric constant, $\varepsilon_r = 2.3$, loss tangent, $tan\delta=0.002$, and size 10 cm×10 cm×1.57 mm. The width of the 50 Ω microstrip line is $w=4.66$ mm. The DR, made from Zinc Titanate material has $\varepsilon_r = 24$, $tan\delta=0.002$, diameter, $2a= 19.43$ mm and height, $h = 7.3$ mm. The sensor works on the principle that when the DR is displaced between the strips, the magnitude of transmission coefficient at resonance, $|S_{21}|_{max}$ varies. This variation of $|S_{21}|_{max}$ can be related to the resonator and circuit parameters by equation (1) [39]

$$S_{21} = \frac{2\sqrt{k_1 k_2}}{1 + k_1 + k_2 + j2Q_0\delta} \quad (1)$$

Where $Q_0$ is the unloaded quality factor, $\delta$ is fractional frequency change ($\delta = \frac{f-f_0}{f_0}$), and $k$'s are the coupling coefficients with respect to the input and output lines ($k \geq 0$). At resonance i.e., $f=f_0$, $\delta=0$ causing $S_{21}=|S_{21}|_{max}$ vary only with the $k$'s as seen from (1). The $k$'s in turn vary with the proximity of the DR to the microstrip lines as it happens in a typical resonator-coupled transmission line circuit.

For the simulation model in Fig.1(a), the plots of normalized $|S_{21}|_{max}$ vs the horizontal ($dx$) and vertical ($dy$) displacements at the DR's resonant frequency of 3.69 GHz are shown in Fig.1(b). As Fig.1(b) reveals, this sensor has a horizontal detection range of 18 mm and a vertical range of 5 mm, with exponential sensitivity responses. Corresponding average sensitivities are 1.25 dB/mm and 3.9 dB/mm in the $x$ and $y$ directions respectively. This non-linear nature combined with differing sensitivities and ranges in the orthogonal directions are the main drawbacks of this sensor with respect to 2D sensing. This anomaly is caused both by the 1D symmetry of the structure and the cross-coupling (unwanted energy exchange quantified by the $|S_{21}|$ in the absence of the DR) between the parallel microstrip lines. This limits the noise floor of the $|S_{21}|_{max}$ data. Hence a modified sensor design is shown in Fig.2(a), in which the microstrip lines are aligned in the orthogonal directions.

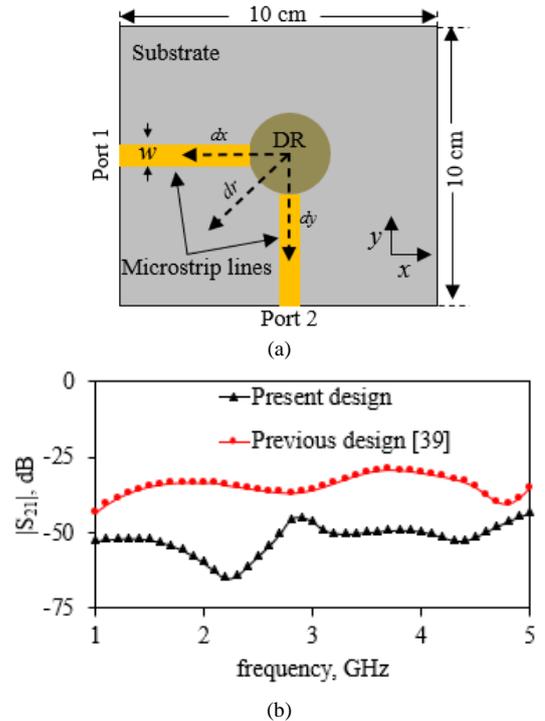

Fig. 1: (a) Schematic diagram (top view) of the DR coupled to parallel microstrip lines [39] (b) Normalized $|S_{21}|_{max}$ response with $dx$ and $dy$

Fig.2: (a) Schematic diagram (top view) of the DR coupled to orthogonal microstrip lines, (b) Cross-coupling levels for the parallel and orthogonal microstrip line circuits of Fig.1 and Fig.2

Apart from this alignment, every other aspect of the circuit is the same as that in Fig.1(a). Thus, the required 2D structural symmetry is achieved (Fig.2 (a)). In addition, the cross-coupling is also significantly reduced compared to the parallel strip arrangement [39] as plotted in Fig. 2 (b). As shown in Fig.2(a), the latter design enables the DR displacement to be detected in the 2D space enclosed by the strips. Also in figure, three representative paths, namely the horizontal ($dx$ @ $dy$=0), vertical ($dy$ @ $dx$=0), and diagonal ($dr$@ $dx=dy$) paths are marked to analyze the sensitivities.

### III. SENSITIVITY ANALYSIS OF THE 2D SENSOR

#### A. Horizontal and vertical sensitivities

ANSYS HFSS [46] software is used to obtain the transmission ($|S_{21}|$) spectrum of the circuit (Fig.2(a)) for varying DR positions ($dx$ and $dy$). Maximum coupling between the DR and strips (indicated by $|S_{21}|_{max}$) occurs when the DR is closer to the open ends of both the strips, i.e. at $dx=dy$=7 mm from the substrate center, to be precise. This position is designated as the reference position, i.e., $dx$ (or $dy$) =0 mm. The corresponding resonant frequency is identified as 3.88 GHz for which the electric and magnetic field distributions in the DR are shown in Fig.3. From these, the electromagnetic mode excited in the DR can be identified as $TE_{011+\delta}$-like mode similar to that in [39]. The $|S_{21}|$ spectrum versus DR position furnished in Fig. 4. (a) reveals that the magnitude of the resonance peak decreases monotonically with an increase in $dx$ (or $dy$). Thus $|S_{21}|_{max}$ provides the magnitude of displacement while the sense of magnitude change (increase or decrease) indicates the displacement direction. Negative displacements are not included in the discussion as it was found that $|S_{21}|_{max}$ decreases much faster causing early saturation of the response.

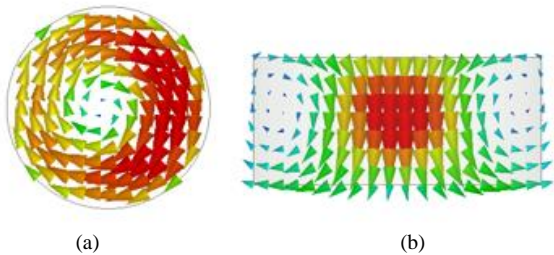

(a)          (b)

Fig. 3: (a) Top view of electric field intensity ($E$) and (b) Side view of magnetic field intensity ($H$) distributions of the $TE_{011+\delta}$ like mode of the DR

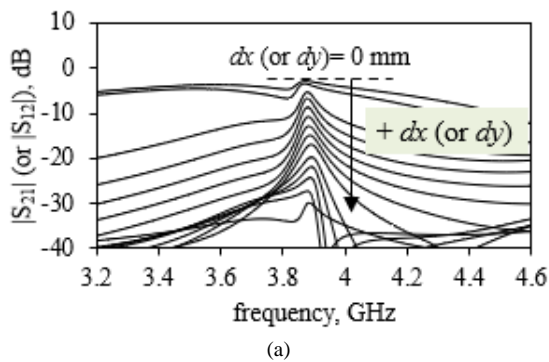

(a)

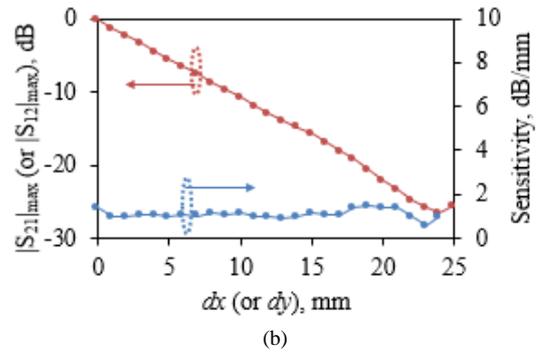

(b)

Fig.4: (a) Simulated $|S_{21}|$ versus frequency for the DR sensor in $dx$ and $dy$ directions (only selected curves are shown for legibility) (b) Simulated response of normalized $|S_{21}|_{max}$ with displacement and corresponding sensitivity plots at $f_0$= 3.88 GHz

Further analysis of the $|S_{21}|$ spectrum of Fg.4(a) reveals that for smaller displacements, the spectrum has a band-pass response (low-Q) superimposed on an all-pass response (high-Q). The band-pass response is indicative of the weakly excited DR mode ($TE_{011+\delta}$–like) while the all-pass response corresponds to the coupled TEM-like transmission line mode of the strips. The TEM-like mode coupling is enhanced by the close proximity of the high permittivity DR to the strips' open ends. For larger displacements, the DR mode dominates the strip mode as justified by the band-pass response of Fig.4(a). Still, one can clearly detect a distinct resonant peak at a constant frequency in the $|S_{21}|$ spectrum around 3.88 GHz whose magnitude decreases monotonically with the increase in displacement. The normalized $|S_{21}|_{max}$ versus displacement and its derivative representing the sensitivity are shown in Fig.4 (b). It clearly shows the enhancement in dynamic range and linearity compared to Fig.1(b). The average sensitivity calculated from the sensitivity curve is 1.01 dB/mm ($dx$ or $dy$) over a dynamic range of 24 mm.

#### B. Diagonal sensitivity ($dr$)

The diagonal movement of the DR results in peak $|S_{21}|_{max}$ at $dr$=10 mm from the substrate center (at $dx=dy$=7 mm). The corresponding normalized $|S_{21}|$ spectrum is furnished in Fig.5 (a). From the figure, the resonant frequency is found as 3.88 GHz, as expected. The corresponding normalized sensitivity curve is shown in Fig.5 (b). The sensitivity curve is quasi-linear with an average sensitivity of 1.25 dB/mm.

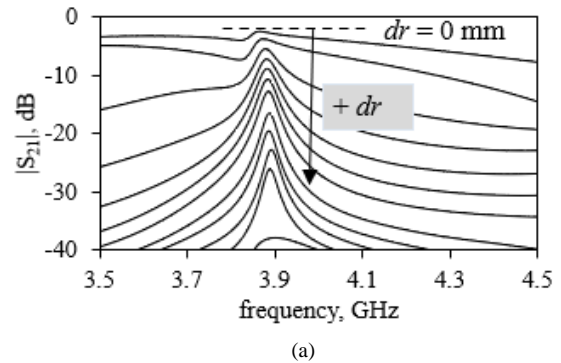

(a)

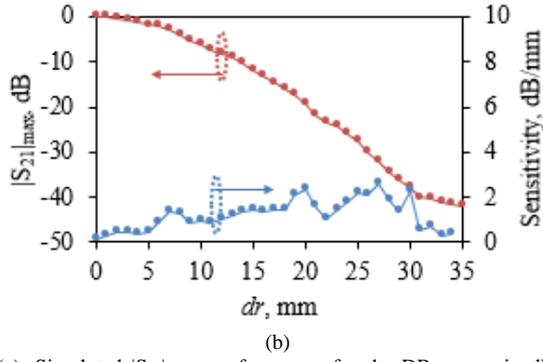

(b)

Fig.5(a): Simulated $|S_{21}|$ versus frequency for the DR sensor in diagonal (*dr*) direction (only selected curves are shown for legibility), (b) Simulated response of normalized $|S_{21}|_{max}$ with displacement and corresponding sensitivity plots at $f_0$= 3.88 GHz

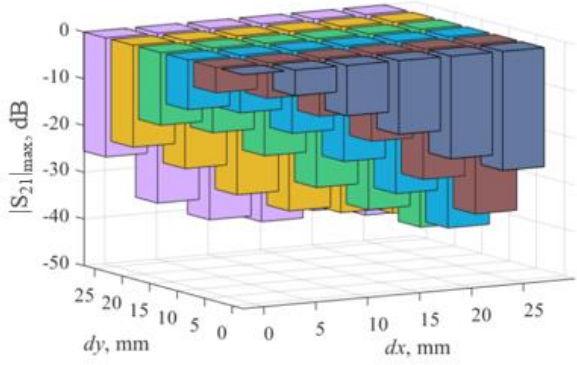

(a)

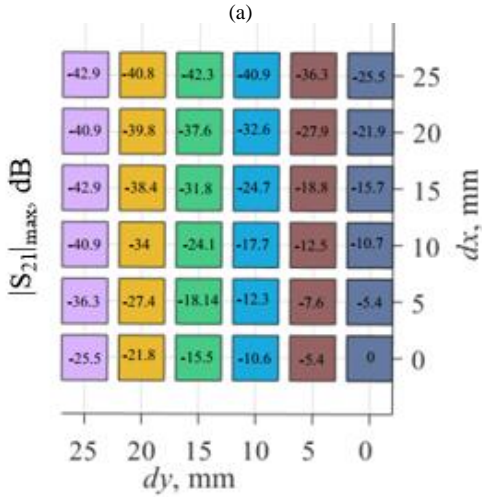

(b)

Fig. 6: Simulated $|S_{21}|_{max}$ response of the proposed sensor in full 2D space (*dx,dy*) (a) Perspective view (b) Top view

Now the simulated $|S_{21}|_{max}$ response over the full 2D space (*dx,dy*) bounded by the strips is furnished in Fig.6. Note that the $|S_{21}|_{max}$ values are normalized with respect to the peak $|S_{21}|_{max}$ that occurs at the reference position i.e., (*dx,dy*) = (0,0). Using these data (Fig.6(b)), already reported 1D curves (Fig.4(b) and 5(b)) can be verified.

## IV. PROTOTYPE AND EXPERIMENTAL RESULTS

The photograph of the fabricated sensor prototype is shown in Fig.7. It is first characterized with a standard 50 Ω vector network analyzer (Keysight N9928A VNA) to record the complete $|S_{21}|$ spectrum as done in the simulation. Full 2-port calibration of the VNA with port extensions (coaxial cables) is performed using the 85052D Cal kit over the frequency range of 3–4.25 GHz with 801 frequency points to achieve high accuracy. After calibration, the uncertainty of $|S_{21}|_{max}$ is reduced from −1.12 dB to −0.03 dB at the resonant frequency. To demonstrate the cost-benefit of this fixed-frequency sensor, a simple RF signal generator & power meter combination (both 50 Ω) in place of the VNA is used [39],[40]. Fig.8 shows both of these measurement approaches.

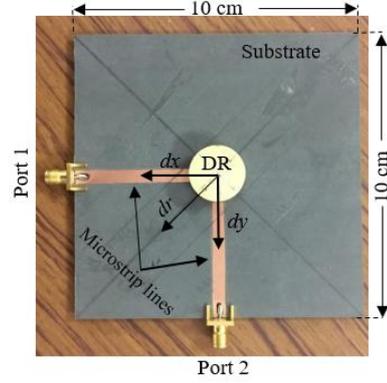

Fig.7: Photograph of the fabricated sensor prototype (parameters are the same as in Fig.2).

Measured $|S_{21}|_{max}$ response from the VNA is shown in Fig.9, for both the *dx@ dy=0* (or *dy@dx=0*) and *dr* (*dx=dy*) displacements, which match very well with the simulated curves of Fig.4. (a) and 5. (a) except for a small decrease in the resonant frequency, i.e., 3.88 GHz vs 3.67 GHz. Measured $|S_{21}|_{max}$ vs displacement curves at 3.67 GHz are shown in Fig.10 (a), (b), which is also in good agreement with the simulation. Note that the DR is displaced manually between the strips with respect to rulers affixed to the substrate, hence, to reduce the positional uncertainty, the above measurements are repeated five times and the average values are reported. Average sensitivity in the *dx* (or *dy*) direction is 1.04 dB/mm identically for both methods, while it is 1.01 dB/mm in simulation. In the *dr* direction, the VNA measurement gives 1.14 dB/mm, whereas it is 1.2 dB/mm for signal generator & power meter measurement. In the simulation, the same is 1.25 dB/mm. The dynamic range in *dx* (or *dy*) directions is 23 mm while it is 30 mm in *dr* direction, in all the cases.

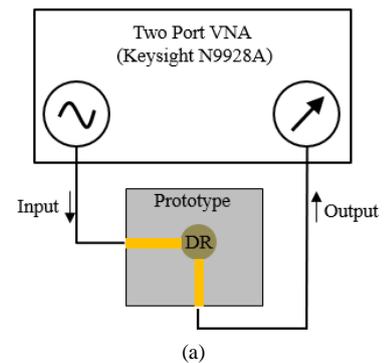

(a)

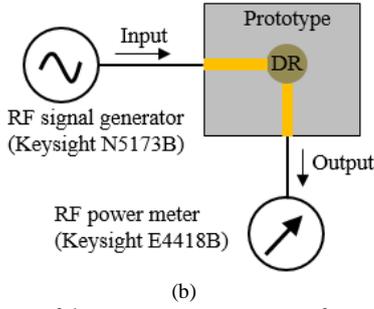

(b)

Fig. 8: Block diagram of the two measurement setups for sensor validation using (a) VNA (b) Signal generator & powermeter

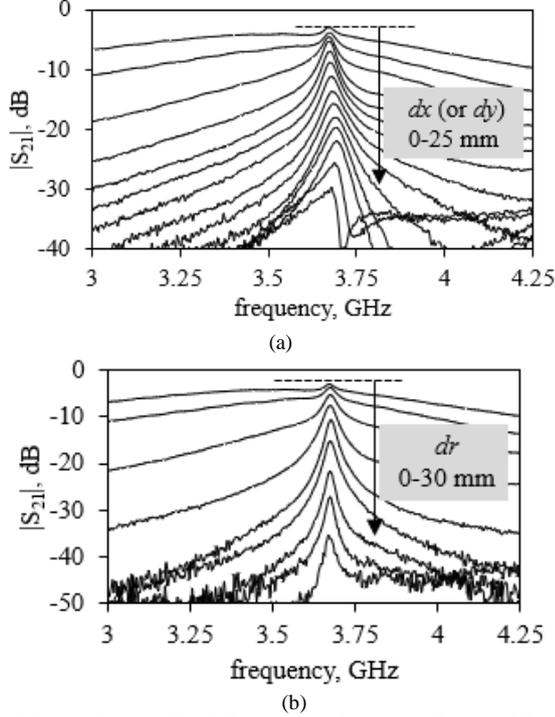

Fig.9: Measured normalized $|S_{21}|$ versus frequency for the DR sensor prototype at 3.67 GHz for displacement in the (a) Orthogonal directions $dx$ (or $dy$) when $dy$ (or $dx$)=0 (b) Diagonal directions ($dr$) when $dx=dy$ (Only selected curves are shown for legibility)

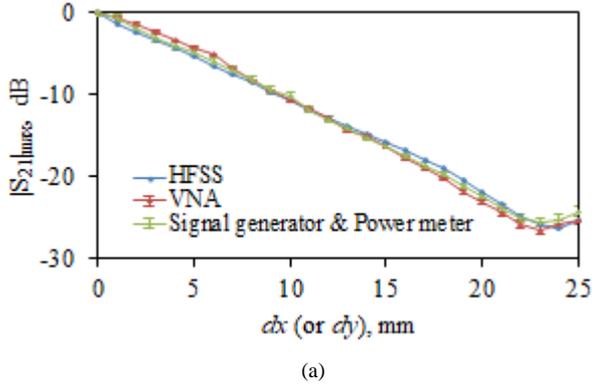

(a)

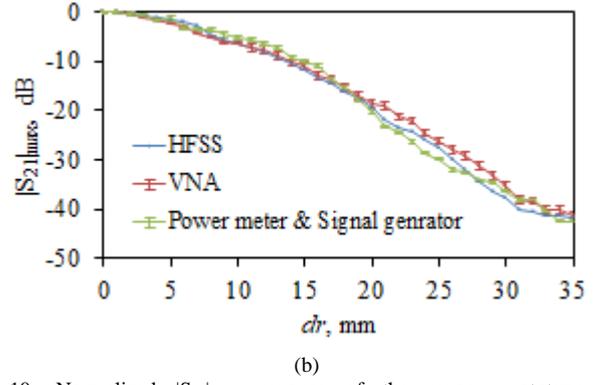

(b)

Fig.10: Normalized $|S_{21}|_{max}$ response of the sensor prototype for displacement in the (a) Orthogonal directions $dx$ (or $dy$) when $dy$ (or $dx$) =0 (b) Diagonal directions ($dr$) (HFSS at 3.88 GHz, others at 3.67 GHz, Error bars for measured points are also shown)

## V. REGRESSION AND ERROR ANALYSIS

As the measurement is performed with two separate methods, for convenience of reference, the VNA method is termed as method 1 and the Signal generator & power meter method is termed as method 2. A linear regression fit of the measured $|S_{21}|_{max}$ data yields the R-squared values (coefficient of determination) of 0.9973 for method 1 and 0.9986 for method 2 for the $dx$ or $dy$ displacement. A polynomial regression fit gives R-squared values of 0.9992 for method 1 and 0.9897 for method 2 for the $dr$ displacements, as shown in Fig.11. (a) and (b).

For the proposed sensor, one or more of the tolerances of fabrication, assembly and proximity effects lead to measurement errors. As it was observed before, the measured resonant frequency is 210 MHz lower than that of the simulation, but the deviation between measured and simulated sensitivity curves (Fig.10) is very small. This is the unique advantage of the proposed fixed frequency-variable magnitude sensor, compared to variable-frequency sensors. The root mean squared (RMS) error between the measured and curve fit values (estimated) over the dynamic range of 0–23 mm for 2D displacement (Fig.10(a)), and 0–30 mm for diagonal displacement (Fig.10(b)) are calculated. For the 2D ($dx$ or $dy$) displacement, the errors are 0.97 dB for method 1 and 0.52 dB for method 2. For the diagonal ($dr$) direction, the error is 1.47 dB for method 1 and 1.31 dB for method 2. With the use of micrometer actuators [9], [21] to displace the DR between the strips, the above errors may be greatly reduced. The performance metrics of the proposed 2D displacement sensor are summarized in Table. I.

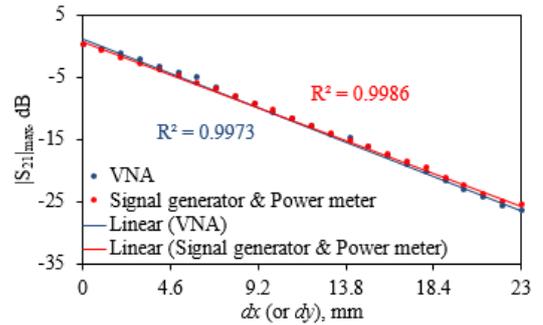

(a)

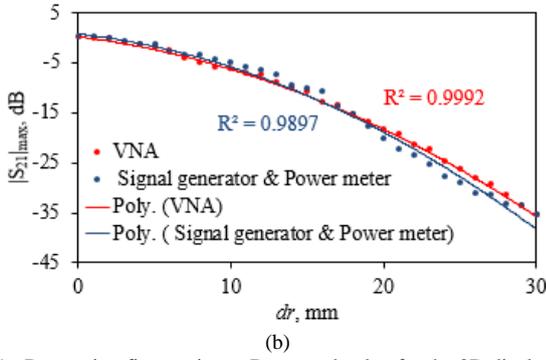

(b)

Fig.11: Regression fit to estimate R-squared value for the 2D displacement sensor in (a) *dx* (or *dy*) direction, (b) *dr* direction

Table. I
Performance metrics of the proposed displacement sensor extracted from VNA measurements and simulations

| Performance metric | *dx* (or *dy*) | *dr* |
|---|---|---|
| Sensitivity (dB/mm) | 1.04 | 1.14 |
| Dynamic range (mm) | 23 | 30 |
| R-squared Value | 0.9973 | 0.9992 |
| RMS error (dB) | 0.97 | 1.47 |

## VI. MAPPING OF FULL 2D SPACE

The simulated 2D plot of $|S_{21}|_{max}$ vs space (Fig.6) shows that the transmission coefficient, $S_{21}$ exhibits diagonal symmetry. This is also indicative of the reciprocity of the structure i.e, $S_{21}=S_{12}$. Hence this parameter is not sufficient to differentiate between the upper and lower triangular regions of the substrate space. However, depending on the position of the DR about the diagonal, the DR's proximity with the respective feed varies, which in turn changes the input impedance and reflection coefficients at both the ports by varying degrees. For example, simulated $|S_{11}|$ and $|S_{22}|$ versus frequency for the *dx* movement @ *dy=0* are shown in Fig.12(a) and (b) respectively. Corresponding $|S_{11}|_{min}$ and $|S_{22}|_{min}$ at 3.88 GHz versus *dx* are plotted in Fig. 12(c). The figures show that for $dx \leq 5$ mm, $|S_{11}|_{min}$ decreases from −7 to −32 dB, and for $dx > 5$ mm, it increases towards 0 dB exhibiting marginal variation with *dx*, in a non-monotonic fashion. On the other hand, the $|S_{22}|_{min}$ variation is limited to a smaller range of −7 to 0 dB only over the entire *dx* range of 25 mm. It is obvious that the above behavior of $|S_{11}|$ and $|S_{22}|$ will reverse for *dy* movement @*dx=0*. Also, for the *dr* movement i.e., *dx=dy*, $S_{11}=S_{22}$ (due to symmetry). This suggests that the differential reflection parameter $\Delta S_{ii}=|S_{11}|_{min}-|S_{22}|_{min}$ would be useful in dictating the one-to-one mapping between the $|S_{21}|_{max}$ and the 2D space. Simulated and measured $\Delta S_{ii}$ variations for three specific paths (*dx, dy,* and *dr*) are plotted in Fig.13. It can be readily made out from the figure that if $\Delta S_{ii} = 0$, the DR is centered at the diagonal and if $\Delta S_{ii} < 0$ the DR is on the upper (lower) triangular space, and if $\Delta S_{ii} > 0$ DR is in the lower (upper) triangular space.

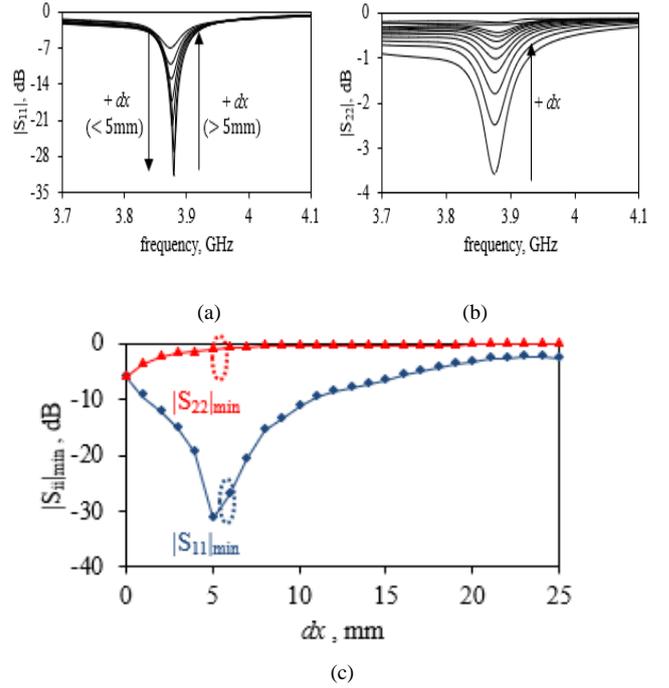

Fig.12: Simulated reflection responses for *dx@dy=0* displacement (a) $|S_{11}|$ vs frequency (b) $|S_{22}|$ vs frequency (only selected curves are shown for legibility) (c) Corresponding normalized $|S_{ii}|_{min}$ response at $f_0$= 3.88 GHz

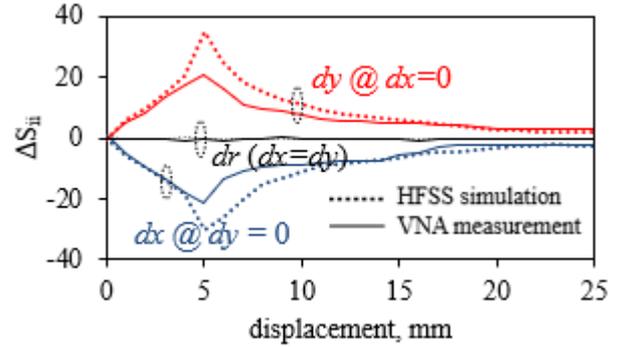

Fig.13: Simulated vs measured differential parameter $\Delta S_{ii} = |S_{11}|_{min}-|S_{22}|_{min}$ vs *dx*, *dy* and *dr* for direction detection

To further verify the mapping, measured $|S_{21}|_{max}$, and $\Delta S_{ii}$ over the full 2D space of the substrate are furnished in Fig.14(a) and (b) respectively. These results are in good agreement with the conclusions made from the 1D positioning results discussed above. The above 2D positioning data can be used to train an Artificial Neural Network (ANN)-based machine learning algorithm [47] for accurate inverse mapping of the S-parameter data to the corresponding DR position. Although the above 2D positioning test used only the VNA measurements, a low-cost complete S-parameter measurement setup similar to Fig.8 can also be used for cost-effective measurement.

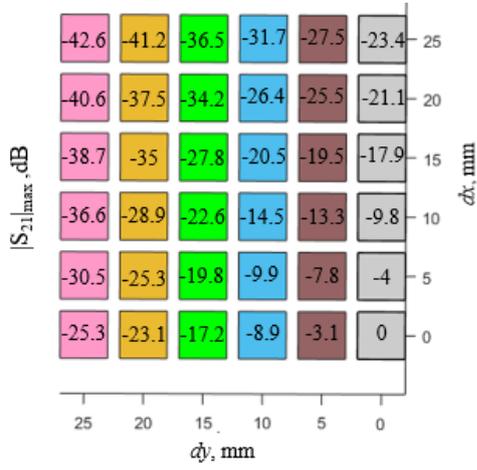

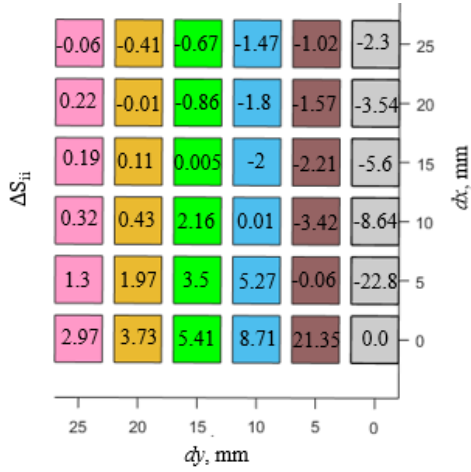

Fig. 14: VNA measured data for full 2D space of the substrate
(a) $|S_{21}|_{max}$ (b) $\Delta S_{ii}$

## VII. DR PARAMETERS AND SENSOR PERFORMANCE

To understand how the present 2D sensor performance (operating frequency, average sensitivity, and dynamic range) depends on the DR parameters (dielectric constant-$\varepsilon_r$, and aspect ratio-$a/h$ for a given diameter-$2a$), a detailed parametric analysis by using HFSS is presented in this section. The selected values of each parameter ($\varepsilon_r$: 16, 20, 24, 30, and $a/h$: 0.5, 0.7, 1, 1.3, 1.5) are based on the specification chart from commercial DR material supplier [48]. The DR diameter is retained same as the nominal value of $2a$=19.43 mm. As the sensitivity in the diagonal direction is indicative of the horizontal and vertical sensitivities, this parametric study considers the diagonal sensitivity ($dr$) only Results of the parametric sweep are plotted in Fig.15.

At first, the effect of the DR parameters on the resonant frequency is investigated in Fig. 15(a). It can be observed that the resonant frequency varies inversely with the dielectric constant, while it varies directly with the aspect ratio of the DR. This feature can be used to design the sensor frequency to avoid electromagnetic interference (EMI) from nearby wireless devices and services. Next, the displacement sensitivity dependence on the DR parameters is considered in Fig. 15(b). The sensitivity is observed to be inversely proportional to the dielectric constant while it is directly proportional to the aspect ratio. Thus, it can be concluded that the sensitivity can be increased by increasing the DR's resonant frequency through increasing its aspect ratio (i.e., a low profile DR) and/or decreasing the dielectric constant. Lastly, the dynamic range variation is shown in Fig. 15(c) which implies that it is less dependent on the DR's parameters but rather on the coupling circuit design.

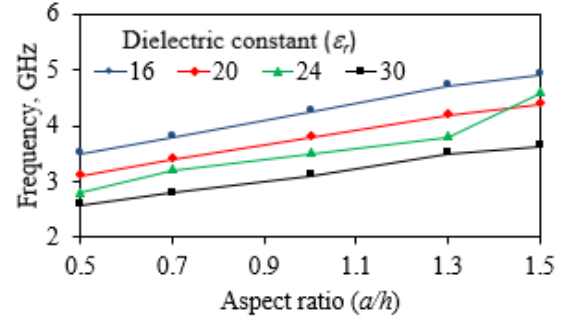

(a)

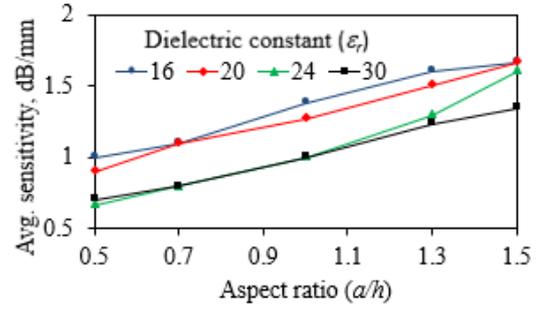

(b)

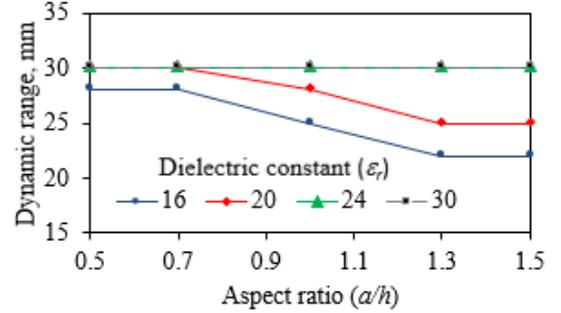

(c)

Fig.15: Effect of DR parameter variation on the sensor performance (a) Frequency (b) Average sensitivity (c) Dynamic range

Table. II
Performance comparison of the proposed sensor with previously published 2D displacement sensors

| Reference | Sensing element | $f_0$ (GHz) | Sensor principle | Sensitivity | Dynamic range (mm) | 2D positioning test |
|---|---|---|---|---|---|---|
| [18] | BC-SRR | 2.5 | Variable frequency | 0.10 MHz/mm | 3 | NO |
| [19] | CSRR | 2.4 | Variable frequency | 50-110 MHz/mm | 4 | NO |
| [21] | DGS | 3-5 6-10 | Variable frequency | $dx$: 0.41, $dy$:1.2 MHz/ mm | 3 | NO |
| [22] | FSS | 11.37 13.7 | Variable frequency | $dx$:280, $dy$: 372 MHz/ mm | 5 | NO |
| [24] | SIR | 1.98 | Variable frequency | $dx$: 520 MHz/mm $dy$ : NA | 7 | NO |
| [20] | SRR | 4.25 | Fixed frequency-Variable magnitude | $dx$ & $dy$: 19 dB/mm | 0.8 | NO |
| This work | DR | 3.67 | Fixed frequency-Variable magnitude | $dx$ & $dy$: 1.04 dB/mm $dr$: 1.14 dB/mm | $dx$ & $dy$: 23 $dr$: 30 | YES |

## VIII. DISCUSSION AND CONCLUSION

In Table. II, the proposed sensor design and performance is compared with those of existing 2D sensors. The present sensor's performance is experimentally validated in the 2D space, unlike others in the list which report only 1D results. Sensors in [18]−[23] use multiple resonators whereas the proposed sensor uses a single resonator. Air-gap sensitivity which is the major source of frequency and magnitude errors in [18] and [20], is avoided in the present sensor by keeping the DR in contact with the substrate. Also, the present sensor offers the highest dynamic range of all sensors. The dynamic range can further be enhanced by using longer strips (i.e., larger substrate) as observed in simulations (not shown).

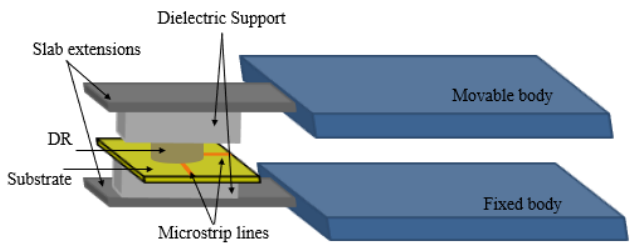

Fig.16. Illustration of a practical application scenario of the proposed sensor module

Fig.16 illustrates a practical application scenario where the proposed 2D sensor may be employed. It comprises two bodies, one stationary and the other movable. As shown, the substrate of the sensor forms part of the stationary body and the DR forms part of the movable body. Proximity effects of external entities like support structures, screws, covers, etc. are expected in such a practical environment. However, due to the fixed frequency-variable magnitude operation, the above perturbations will have negligible effects on the sensor performance. As the proposed sensor is compact, it easily integrated, accurate, especially in harsh environments, and provides a wide dynamic range, it is suggested as a useful candidate for industrial applications that require the detection and correction of alignment and positional errors between two bodies. Further investigations in this direction are left for future work.

ACKNOWLEDGMENT

The authors are thankful to Mr. Mahesh Chandra Saini (Technical assistant, EEE Department, BITS Pilani) for his assistance in prototype fabrication.